\documentclass[journal,draftclsnofoot,onecolumn,12pt]{IEEEtran}

\newif\ifarxiv
\arxivtrue 

\newif\ifonecolumn
\onecolumntrue 

\IEEEoverridecommandlockouts

\usepackage{cite}
\usepackage{amsmath,amssymb,amsfonts}
\usepackage{float}
\usepackage{caption}
\usepackage{graphicx}
\usepackage{textcomp}
\usepackage{algorithm}
\usepackage[left=0.625in,right=0.625in,bottom=0.95in,top=0.75in]{geometry}
\usepackage{amsthm}
\usepackage{algpseudocode}
\usepackage{verbatim}
\usepackage{paralist}
\usepackage{array}
\newcolumntype{?}{!{\vrule width 1.1pt}}
\usepackage{tablefootnote}

\newtheorem{proposition}[]{Proposition}

\newtheorem{lem}[]{Lemma}

\theoremstyle{remark}

\usepackage{xcolor} 
\definecolor{green}{rgb}{0.0, 0.5, 0.0} 
\allowdisplaybreaks
\usepackage{acro}

\DeclareAcronym{snr}{
  short = SNR,
  long = signal-to-noise ratio,}

  \DeclareAcronym{sar}{
  short = SAR,
  long = synthetic aperture radar,}

\DeclareAcronym{insar}{
  short = InSAR,
  long = interferometric synthetic aperture radar,}

\DeclareAcronym{uav}{
        short = {UAV},
        long = {unmanned aerial vehicle},
        long-plural-form = {unmanned aerial vehicles}
}

\DeclareAcronym{fdma}{
	short = {FDMA},
	long = {frequency-division multiple-access},
}

\DeclareAcronym{3d}{
        short = {3D},
        long = {three-dimensional},
}

\DeclareAcronym{2d}{
        short = {2D},
        long = {two-dimensional},
}

\DeclareAcronym{dem}{
        short = {DEM},
        long = {digital elevation model},
}
\DeclareAcronym{gs}{
        short = {GS},
        long = {ground station},
        long-plural-form = {ground stations}
}

\DeclareAcronym{los}{
        short = {LOS},
        long = {line-of-sight},
}

\DeclareAcronym{sca}{
        short = {SCA},
        long = {successive convex approximation},
}

\DeclareAcronym{ao}{
        short = {AO},
        long = {alternating optimization},
}

\DeclareAcronym{nesz}{
        short = {NESZ},
        long = {noise equivalent sigma zero},
}

\DeclareAcronym{wrt}{
        short = {w.r.t.},
        long = {with respect to },
}
\DeclareAcronym{rhs}{
        short = {r.h.s},
        long = {right-hand side },
}

\DeclareAcronym{lhs}{
        short = {l.h.s},
        long = {left-hand side },
}
\DeclareAcronym{hoa}{
        short = {HoA},
        long = {height of ambiguity},
}

\begin{document}

\title{UAV Formation Optimization for Communication-assisted InSAR Sensing\\
\thanks{This work was supported in part by the Deutsche Forschungsgemeinschaft (DFG, German Research Foundation) GRK 2680 – Project-ID 437847244.}
}
\author{\IEEEauthorblockN{Mohamed-Amine~Lahmeri\IEEEauthorrefmark{1}, Víctor Mustieles-Pérez\IEEEauthorrefmark{1}\IEEEauthorrefmark{2}, Martin Vossiek\IEEEauthorrefmark{1}, Gerhard Krieger\IEEEauthorrefmark{1}\IEEEauthorrefmark{2}, and
Robert Schober\IEEEauthorrefmark{1}}\\
\IEEEauthorblockA{\IEEEauthorrefmark{1}Friedrich-Alexander-Universit\"at Erlangen-N\"urnberg, Germany\\
\IEEEauthorrefmark{2}German Aerospace Center (DLR),  Microwaves and Radar Institute, Weßling, Germany\\
\vspace{-8mm}}}


\maketitle
\begin{abstract} Interferometric synthetic aperture radar (InSAR) is an increasingly important remote sensing technique that enables  three-dimensional (3D) sensing applications such as the generation of accurate digital elevation models (DEMs).
In this paper, we investigate the joint formation and communication resource allocation optimization for a system comprising two   unmanned aerial vehicles (UAVs) to perform InSAR sensing and to  transfer the acquired data to the ground. To this end, we adopt as sensing performance metrics  the interferometric coherence, i.e., the local correlation  between the two co-registered UAV radar images, and the height of ambiguity (HoA),  which together are a measure for the accuracy with which the InSAR system can estimate the height of ground objects. In addition, an analytical expression for the coverage of the considered InSAR sensing system is derived. Our objective is to maximize the InSAR coverage while satisfying all relevant InSAR-specific sensing and communication performance metrics. To tackle the non-convexity of the formulated optimization problem, we employ alternating optimization (AO) techniques combined with successive convex approximation (SCA). Our simulation results reveal  that the resulting resource allocation algorithm outperforms two benchmark schemes in terms of  InSAR coverage, while satisfying all sensing and real-time communication requirements. Furthermore, we highlight the importance of efficient communication resource allocation in  facilitating real-time sensing and unveil the trade-off between InSAR height estimation accuracy and coverage.
\end{abstract}

\section{Introduction}
The widespread use of \acp{uav} has revolutionized modern technology, impacting fields like remote sensing, communication, and disaster  monitoring  \cite{uav_survey}. Their versatility and cost-effectiveness have made them  an indispensable tool for these diverse applications. Specifically, \acp{uav} excel in remote sensing applications due to their remarkable ability to swiftly acquire high-quality data, be it for ranging and detection or imaging purposes \cite{uav_remote_sensing_survey}. In this context, seamless \ac{uav}-to-\ac{uav} as well as \ac{uav}-to-ground connectivity  enables the timely collection and tranfer of essential information in highly dynamic scenarios. In fact, researchers have made significant strides in developing robust communication architectures that can ensure real-time \ac{uav}-based communication even in challenging environments \cite{uav_communication_survey_slim}.\par

\Ac{insar} is  a well-established remote sensing technique, for which the use of \acp{uav} will open up new application opportunities for the high-resolution observation of small-scale areas and the systematic monitoring of local processes  \cite{insar_introduction}.  
\ac{insar} systems employ two \ac{sar} sensors to illuminate a given area from different angles and by analyzing the phase difference of the two received radar signals, information about the topography and temporal variations of the target area can be derived \cite{coherence1}. For \ac{insar}, the sensing area is defined by the region where the ground footprints of the  two side-looking radar antennas overlap.  Furthermore, conventional sensing performance metrics, such as the detection probability and the false alarm rate, are inadequate for evaluating  the \ac{insar} performance. In fact, the key performance metric for estimating interferometric performance  is  coherence, which is a function of the correlation between the  co-registered master and slave \ac{sar} images \cite{coherence1}. Another relevant performance metric is the \ac{hoa}, which is a proportionality constant between the interferometric phase and the terrain height and is thus related to the sensitivity of the radar to the ground topography \cite{coherence1}. An interesting trade-off in performance arises here; while a large inter-\ac{uav} separation distance leads to  a small \ac{hoa}, improving sensing accuracy, it also leads to a degradation of the  image coherence \cite{tradeoff}. Based on the above discussion, existing results for conventional \ac{uav}-based sensing \cite{robert} are not applicable for \ac{uav}-based \ac{insar} sensing. Moreover, while some preliminary UAV-based InSAR experiments have been reported in \cite{burried,victor}, the optimization-based design of these systems has not been yet considered in the open literature.  \par
In this paper, we present the first optimization framework for communication-assisted \ac{uav}-based bistatic \ac{insar} sensing. Our contributions can be summarized as follows:
\begin{compactitem}
\item We adopt \ac{insar}-specific sensing performance metrics such as  \ac{insar} coverage,  \ac{hoa}, and  interferometric coherence for optimization of InSAR systems. 
\item We formulate and solve a joint formation and communication resource optimization problem  to maximize the \acp{uav}' coverage while guaranteeing the pertinent  sensing  and communication constraints.
\item Our simulations  reveal that, in comparison with two benchmark schemes, a significantly larger  ground  area can be covered with the proposed scheme and highlight interesting \ac{insar}-specific performance trade-offs.
\end{compactitem}

{\em Notations}: 
In this paper, lower-case letters $x$ refer to scalar numbers, while boldface lower-case letters $\mathbf{x}$ denote vectors.  $\{a, ..., b\}$ denotes the set of all integers between $a$ and $b$. $|\cdot|$ denotes the absolute value operator. $\mathbb{R}^{N}$ represents the set of all $N$-dimensional vectors with real-valued entries. For a vector $\mathbf{x}\in\mathbb{R}^{N } $, $||\mathbf{x}||_2$ denotes the Euclidean norm, whereas  $\mathbf{x}^T$ stands for the  transpose of $\mathbf{x}$.  For a real-valued multivariate function $f(\mathbf{x})$, $\nabla_{\mathbf{x}}f(\mathbf{a})$ denotes the gradient vector of  $f$ \ac{wrt} $\mathbf{x}$ evaluated for an arbitrary vector $\mathbf{a}$. For real numbers $a$ and $b$, $\max(a,b)$ and $\min(a,b)$ stand for the maximum and minimum of $a$ and $b$, respectively. For a scalar $x \in \mathbb{R}$, $[x]^+$ refers to $\max(0,x)$. The notation $X_n(\mathbf{x}_1,...,\mathbf{x}_i)$ is an equivalent  notation for $X$ highlighting that $X$ depends on optimization variables $( \mathbf{x}_1,...,\mathbf{x}_i)$ and time slot $n$.

\section{System Model} \label{Sec:SystemModel}

\ifonecolumn
\begin{figure}
    \centerline{\includegraphics[width=3.5in]{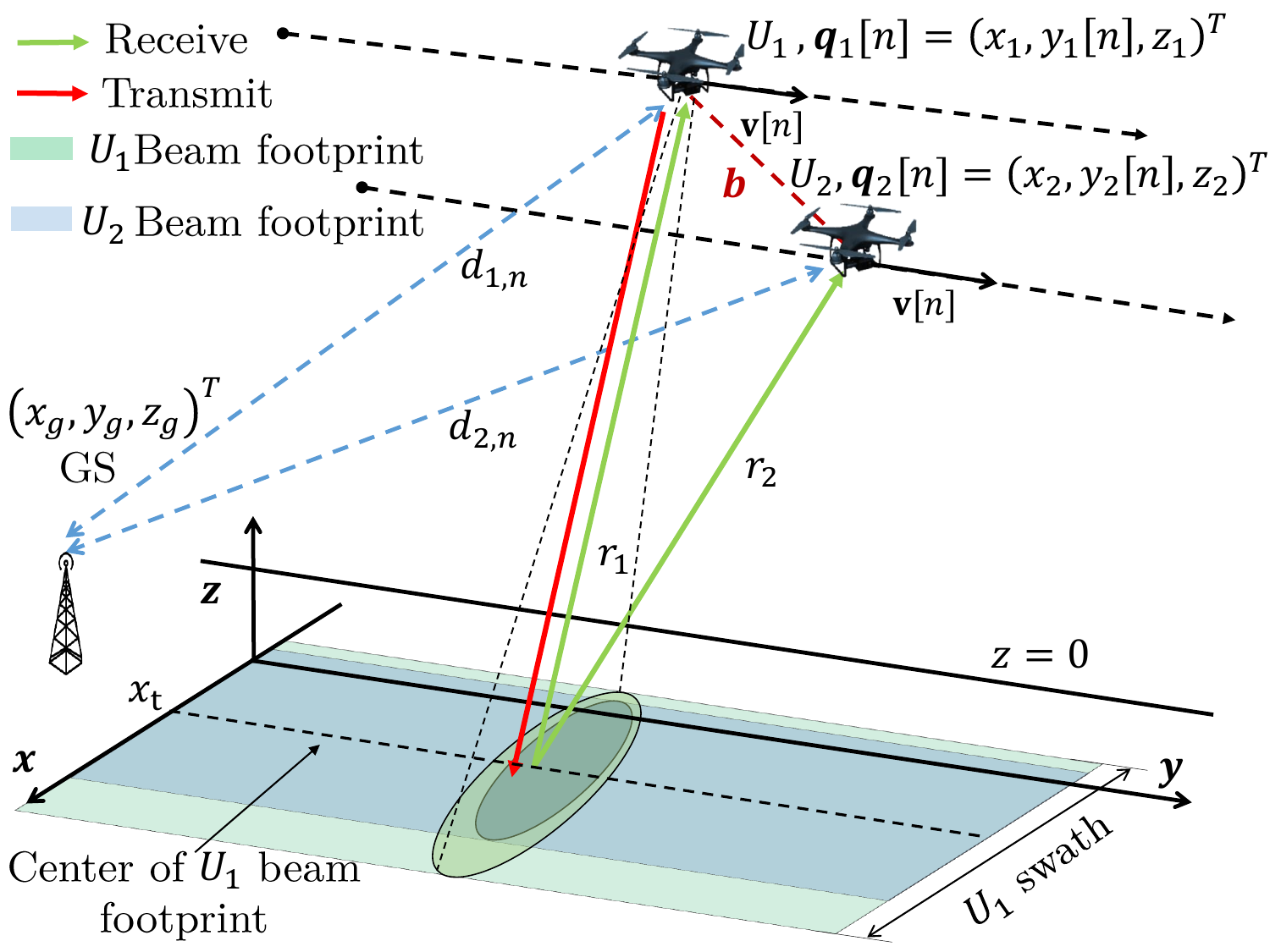}}
    \caption{ InSAR sensing system with two UAV SAR sensors and a GS for real-time data offloading.}
    \label{fig:system-model}
\end{figure}
\else
\begin{figure}
    \centerline{\includegraphics[width=0.9\columnwidth]{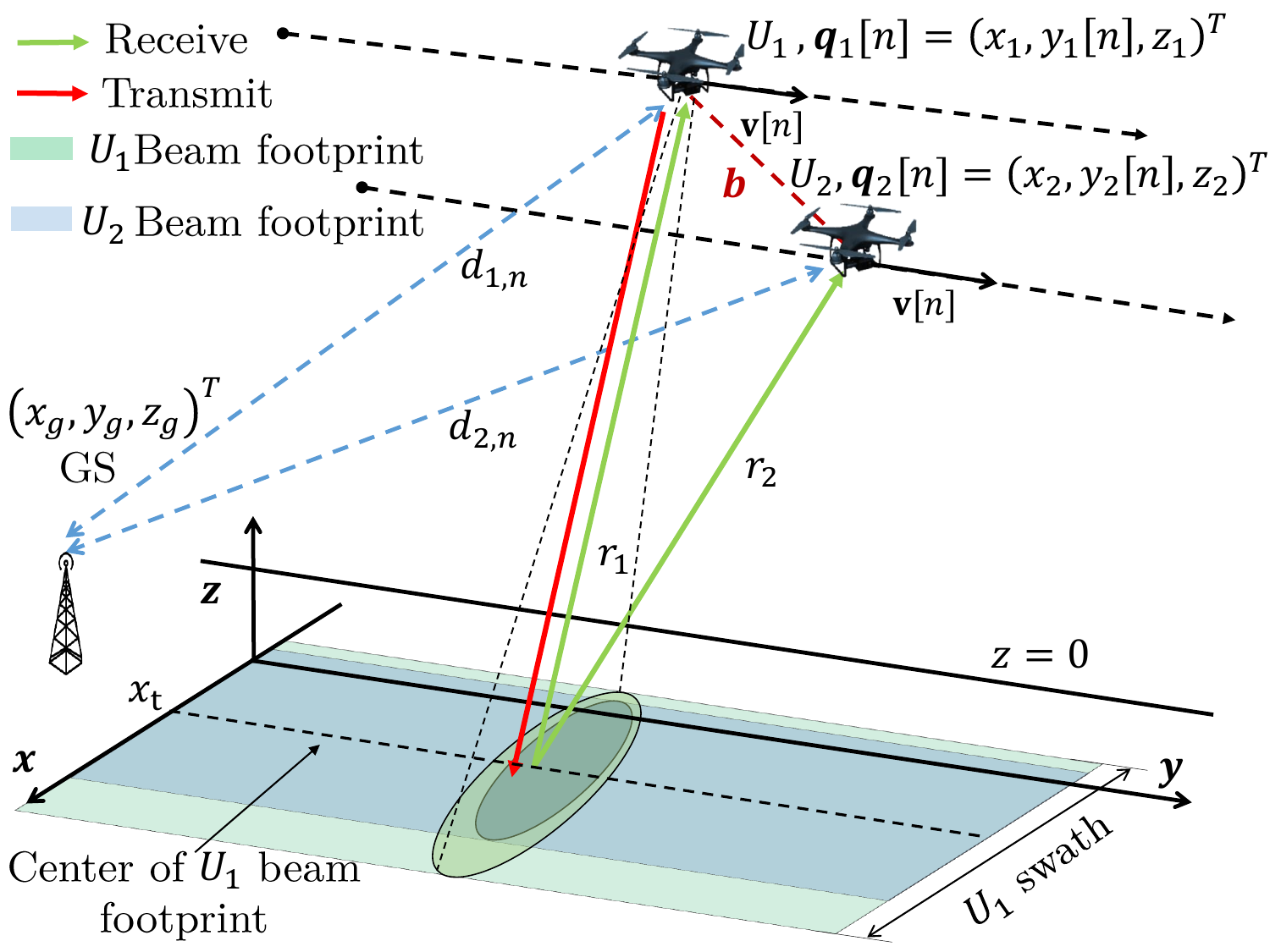}}
    \caption{InSAR sensing system with two UAV SAR sensors and a GS for real-time data offloading.}
    \label{fig:system-model}
\end{figure}
\fi
We consider two rotary-wing \acp{uav}, denoted by $U_1$ and $U_2$, that perform \ac{insar} sensing of a given ground area. $U_1$, serving as the master drone, transmits and receives radar signals, whereas  $U_2$, serving as the slave drone, only receives the echoes.  We adopt a \ac{3d} coordinate system, where the $x$-axis represents the ground range direction, the $y$-axis represents the azimuth direction, and the $z$-axis defines the altitude of the drones above ground, see Figure  \ref{fig:system-model}. We discretize the total mission time $T$ into $N$ uniform time slots $\delta_t$ such that  $T=N \cdot \delta_t$ .
We perform across-track interferometry \cite{insar_introduction}, where both drones are located in the same  $x-z$ plane, also referred to as the across-track plane \cite{insar_introduction}, and  follow a linear trajectory, i.e., the stripmap \ac{sar} imaging mode is employed \cite{fixed_heading}. Therefore, the radar coverage along the $x$-axis, referred to as  swath, is centered \ac{wrt} a line that is parallel to the $y$-axis and passes through point $(x_t,0,0)$, see Figure \ref{fig:system-model}. Usually, multiple swaths are required to cover a large area, therefore maximizing the swath is required \cite{14}. Moreover, for \ac{insar}, to ensure that $U_1$ and $U_2$ are always in the same across-track plane, they fly with the same fixed velocity $\mathbf{v}_y = (v_y[1], ...,v_y[N])^T\in\mathbb{R}^N$ such that in time slot $n$, the velocity vector is given by $\mathbf{v}[n]=(0,v_y[n],0)^T\in\mathbb{R}^3, \forall n$,  \cite{snr_equation}. The location of $U_i$ in time slot $n$ is denoted by $\mathbf{q}_i[n]=(x_i,y[n],z_i)^T$, $ i \in \{1,2\}$, where the $y$-axis position vector $\mathbf{y}=(y[1]=0,y[2],...,y[N])^T\in\mathbb{R}^N$ is given by: 
 \begin{align}
 y[n+1]=y[n]+v_y[n]\delta_t, \forall n \in \{1,N-1\}.
 \end{align}
Hereinafter, as the $y$-axis position is pre-determined, we use the simplified notation $\mathbf{q}_i=(x_i,z_i)^T \in \mathbb{R}^2, \forall i \in \{ 1,2\}$, to denote the position of $U_i$ in the across-track plane. Furthermore, the interferometric baseline, which is the distance between the two \ac{insar} sensors, is given by:
\begin{align}
   b(\mathbf{q}_1,\mathbf{q}_2) = ||\mathbf{q}_2 -\mathbf{q}_1||_2.
\end{align}

\subsection{Bistatic \ac{insar} Coverage}
 Unlike cooperative \ac{uav}-based sensing \cite{detection_probability}, the \ac{insar} coverage is limited to the area in which the beam footprints of $U_1$ and $U_2$ overlap, see Figure \ref{fig:system-model-2}. The coverage problem  for UAV-based sensing is challenging because of its near-range nature which causes the swath width to be limited. 
The usable swath width where the beam footprints of both \acp{uav} overlap can be obtained as follows: 
\begin{align}
  S(\mathbf{q}_1,\mathbf{q}_2 ) &= \Big[ \min\left(x_2+ \tan(\theta_{\mathrm{far}})z_2,x_1 +\tan(\theta_{\mathrm{far}})z_1\right)- \notag \\   &
     \max(x_1 + \tan(\theta_{\mathrm{near}})z_1 ,x_2 + \tan(\theta_{\mathrm{near}}) z_2 )\Big]^+,
\end{align}
where $\theta_{\mathrm{far}}= \theta_d+\frac{\theta_{\mathrm{3dB}}}{2}$, $\theta_{\mathrm{near}}= \theta_d-\frac{\theta_{\mathrm{3dB}}}{2}$,  $\theta_d$ is the depression angle of the \ac{sar} antenna, and $\theta_{\mathrm{3dB}}$ its -3 dB beamwidth in elevation, see Figure \ref{fig:system-model-2}.  Thus, the total area covered by the \ac{insar} radar in time slot $n$ is approximated\footnote{The approximation is due to the elliptical shape of the beam footprint on the ground and becomes negligible for large $N$. } as follows:
\begin{align} 
C_N(\mathbf{q}_1,\mathbf{q}_2 ) =\sum^{N}_{n=1} S(\mathbf{q}_1,\mathbf{q}_2 ) v_y[n] \delta_t.
\end{align}
Now, let $r_1$ and $r_2$ denote the slant range of the radars of $U_1$ and $U_2$, respectively. These slant ranges are given by \cite{insar_introduction}:
\begin{align}
    r_i(\mathbf{q}_i)= \sqrt{ (x_i -x_t)^2 +  (z_i)^2  }, \forall i \in \{1,2\}.
\end{align} To maximize the coverage, the master  \ac{uav} is positioned in the across-track plane such that its \ac{sar} antenna beam footprint on the ground is centered at $x_t$. Furthermore, we assume  $r_2(\mathbf{q}_2)\leq    r_1(\mathbf{q}_1)$ and impose the following condition: 
\begin{align}
x_1=x_t - z_1 \tan(\theta_d). \label{eq:master_drone_placement}
\end{align} 

 \ifonecolumn
\begin{figure}
    \centering
    \includegraphics[width=4in]{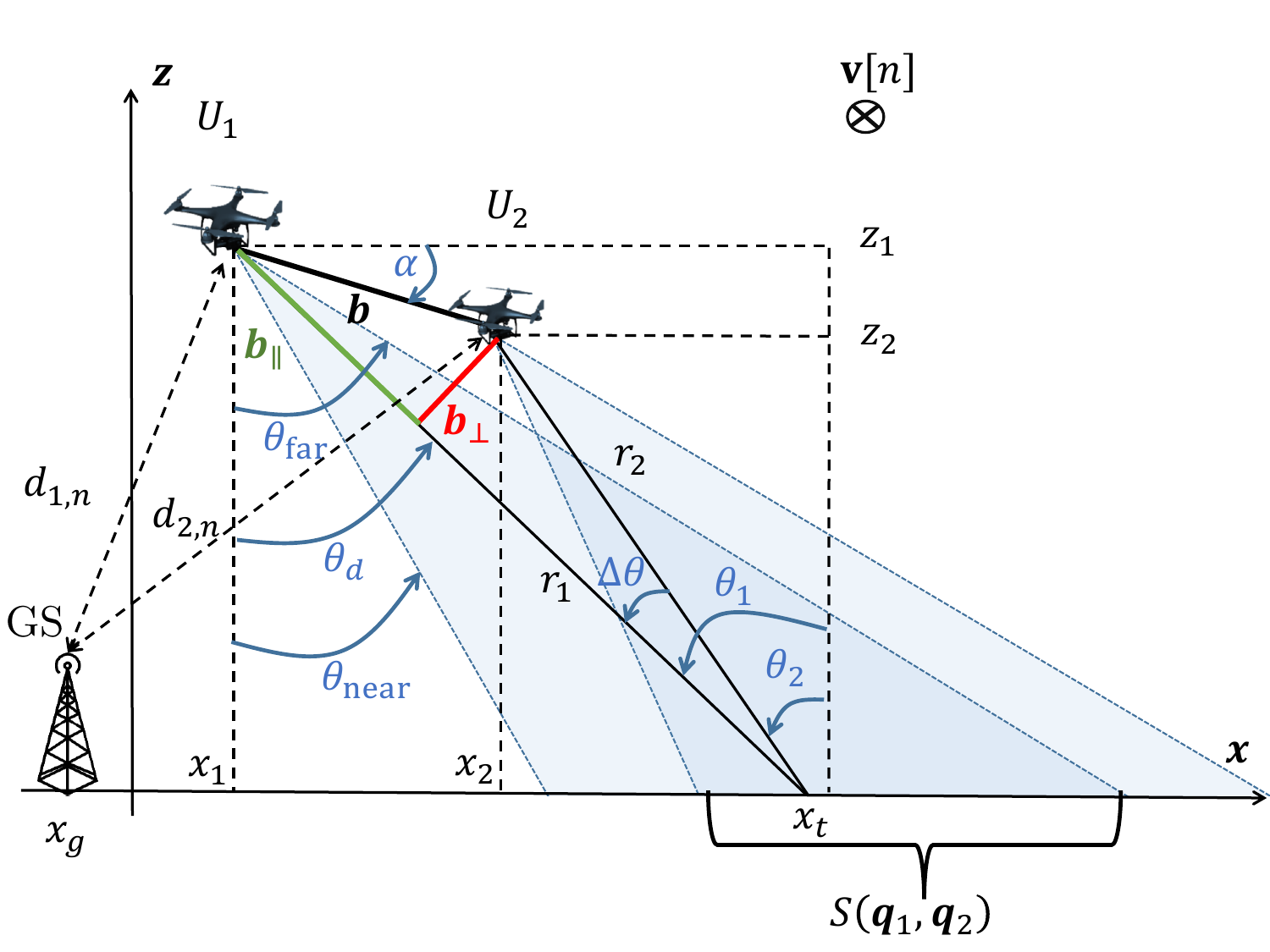}
    \caption{Illustration of the bistatic UAV formation, denoted by $\{\mathbf{q}_1,\mathbf{q}_2\}$, in the across-track plane.}
    \label{fig:system-model-2}
\end{figure}
\else
\begin{figure}
    \centering
    \includegraphics[width=0.9\columnwidth]{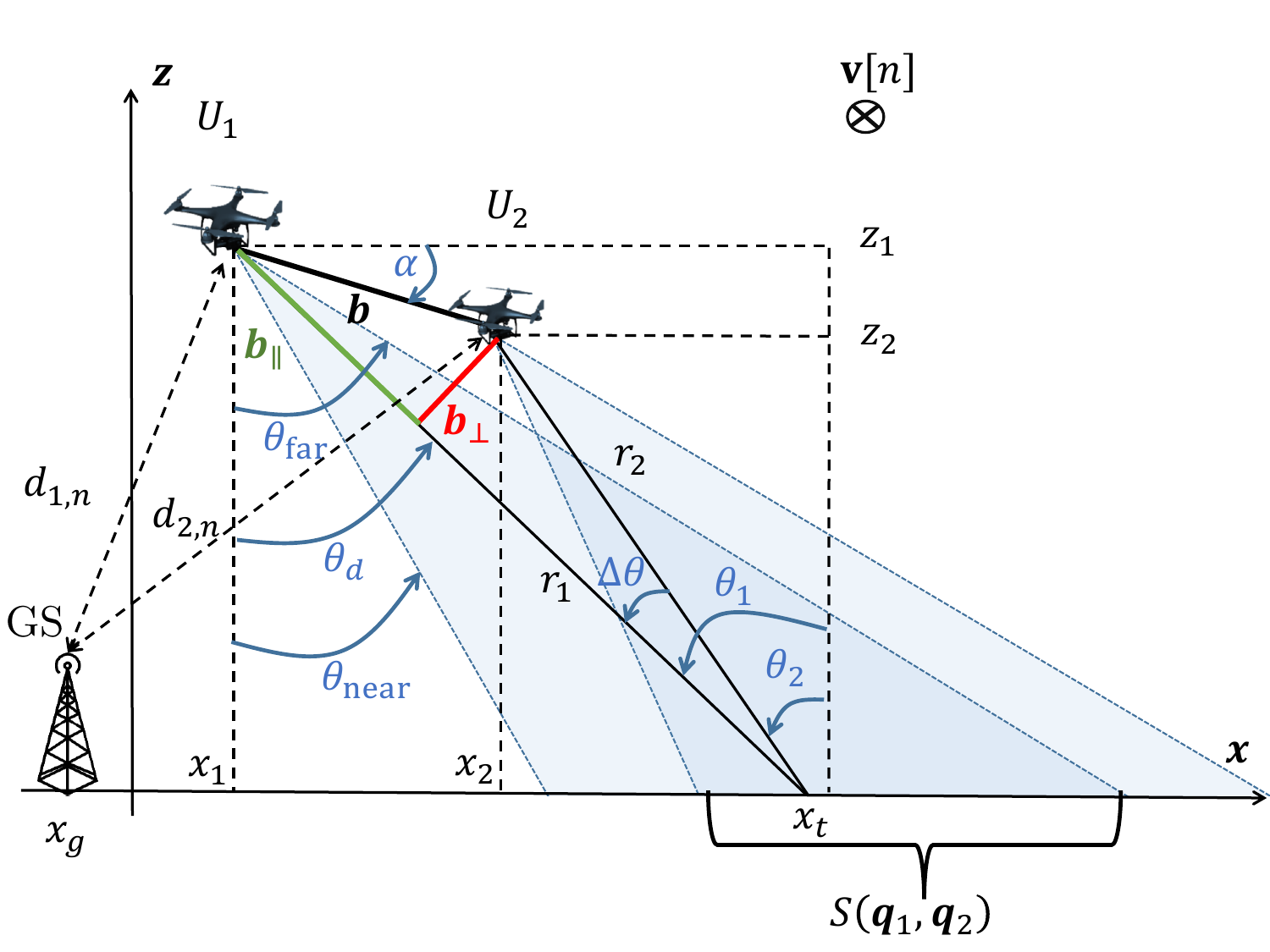}
    \caption{Illustration of the bistatic UAV formation, denoted by $\{\mathbf{q}_1,\mathbf{q}_2\}$, in the across-track plane.}
    \label{fig:system-model-2}
\end{figure}
\fi

\subsection{InSAR Performance}
Next, we introduce the relevant \ac{insar} sensing performance metrics, namely the coherence and \ac{hoa}.
\subsubsection{\Ac{insar} coherence} Important types of decorrelation that affect \ac{insar} coherence, and thereby \ac{insar} performance, are the \ac{snr} decorrelation and baseline decorrelation. In particular,  low \acp{snr}  in \ac{sar} data acquisition result in a loss of coherence between the master and slave \ac{sar} images during processing. In time slot $n$, the resulting \ac{snr} decorrelation is given by  \cite{snr_equation}: 
 \begin{align}
 &\gamma_{\mathrm{SNR},n}(\mathbf{q}_1,\mathbf{q}_2)= \prod_{i \in \{1,2\}} \frac{1}{\sqrt{1+\mathrm{SNR}^{-1}_{i,n}(\mathbf{q}_1,\mathbf{q}_2)}}, \forall n, \label{eq:snr_decorrelation}
 \end{align}
 where $\mathrm{SNR}_{i,n}$ denotes the \ac{snr} achieved by $U_i$ in time slot $n$. In particular, the \ac{snr} achieved by $U_1$ is given by \cite{snr_equation}: 
\begin{equation}
\mathrm{SNR}_{1,n}(\mathbf{q}_1)=\frac{c_n}{r_1^3(\mathbf{q}_1)  \sin(\theta_1) }  , \forall n,
  \end{equation}
  \sloppy where  $\theta_1$ is the angle that $U_1$'s \ac{los}\footnote{The radar \ac{los} is defined by the line connecting the \ac{uav} with the reference point $(x_t,0,0)$.} has with the vertical and $c_n=\frac{\sigma_0 P_t\; G_t\; G_r \; \lambda^3\; c \;\tau_p \;\mathrm{PRF}}{4^4 \pi^3 v_y[n]  k_b T_{\mathrm{sys}} \; B_{\mathrm{Rg}} \; F \; L_{\mathrm{atm}} \; L_{\mathrm{sys}} \; L_{\mathrm{az}}}$. Here, $\sigma_0$ is the normalized backscatter coefficient, $P_t$ is the radar transmit power, $G_t$ is the transmit antenna gain, $G_r$ is the receive antenna gain, $\lambda$ is the wavelength, $c$ is the speed of light, $\tau_p$ is the pulse duration, $\mathrm{ PRF}$ is the pulse repetition frequency, $k_b$ is the Boltzmann constant, $T_{\mathrm{sys}}$ is the receiver temperature, $B_{\mathrm{Rg}}$ is the bandwidth of the radar pulse, $F$ is the noise figure, and  $L_{\mathrm{atm}}$, $L_{\mathrm{ sys}}$, and $L_{\mathrm{ az}}$, represent the atmospheric losses, system losses, and azimuth losses, respectively. The \ac{snr} achieved by the slave \ac{uav} is given by \cite{snr_equation}:
   \begin{equation}
\mathrm{SNR}_{2,n}(\mathbf{q}_1,\mathbf{q}_2)=\frac{c_n}{ r_1^2(\mathbf{q}_1) r_2(\mathbf{q}_2)  \sin(\theta_2(\mathbf{q}_2))}  , \forall n,
\end{equation}
where $\theta_2$ is the angle  that $U_2$'s \ac{los} has  with the vertical. \newline Another relevant type of decorrelation is the baseline decorrelation. It reflects the loss of coherence caused by the acquisition of the two \ac{sar} images in InSAR under  different angles. The baseline decorrelation is  given by\cite{victor}: 
\begin{equation} \gamma_{\mathrm{Rg}}(\mathbf{q}_2)=\frac{(2+B_{p}) \sin\left( \theta_2(\mathbf{q}_2) \right) - (2-B_{ p}) \sin(\theta_1)}{B_p \left(\sin(\theta_1)+\sin\left( \theta_2(\mathbf{q}_2) \right)\right)},\label{eq:baseline_decorrelation}
\end{equation}
where $B_{ p}=\frac{B_{\mathrm{Rg}}}{f_0}$ is the fractional bandwidth and $f_0$ is the radar center frequency. It can be shown that a large interferometric baseline results in high baseline decorrelation and, therefore, degrades the coherence.
\subsubsection{Height of Ambiguity (HoA)} The \ac{hoa} is defined as the height difference which results in a complete $2\pi$ cycle of the interferometric phase \cite{snr_equation}. It is therefore related to the accuracy of the height estimate in the generated \ac{dem}. Similar to the baseline decorrelation,  the \ac{hoa} depends on the \ac{uav} formation and is given by \cite{snr_equation}: 
\begin{align}
    h_{\mathrm{amb}}(\mathbf{q}_1,\mathbf{q}_2)=\frac{\lambda r_1(\mathbf{q}_1) \sin(\theta_1)}{b_{\perp}(\mathbf{q}_1,\mathbf{q}_2)}, \label{eq:HoA}
\end{align}
 where $b_{\bot}$ is the perpendicular baseline, which is the magnitude of the projection of the baseline vector perpendicular to the slant range, see Figure \ref{fig:system-model-2}. The  perpendicular baseline can be obtained as follows:
\begin{gather}
    b_{\bot}(\mathbf{q}_1,\mathbf{q}_2)=
 b(\mathbf{q}_1,\mathbf{q}_2)  \cos\Big(\theta_1- \alpha(\mathbf{q}_1,\mathbf{q}_2)\Big),
\end{gather} 
where $\alpha(\mathbf{q}_1,\mathbf{q}_2)$ is the angle between the interferometric baseline and the horizontal plane, see Figure \ref{fig:system-model-2}. Notice that a large interferometric baseline leads to a small \ac{hoa} value, and therefore, to a better sensitivity to the ground  topography for a given error of the interferometric coherence \cite{coherence1}. 

\subsection{Communication Performance}
We target real-time offloading of the radar data to a communication \ac{gs}, where we adopt \ac{fdma} transmission from the master and slave \acp{uav} to the \ac{gs}. The  instantaneous   transmit power consumed for communication by the master and slave \acp{uav} is given by $\mathbf{P}_{\mathrm{com},1}=(P_{\mathrm{com,1}}[1],...,P_{\mathrm{com,1}}[N])^T \in \mathbb{R}^N$ and  $\mathbf{P}_{\mathrm{com,2}}=(P_{\mathrm{com,2}}[1],...,P_{\mathrm{com,2}}[N])^T \in \mathbb{R}^N$, respectively.
We denote the location of the \ac{gs} by $\mathbf{g}= (x_g, y_g, z_g)^T \in \mathbb{R}^3$ and the distance from $U_i$ to the \ac{gs} by    $d_{i,n}(\mathbf{q}_i) = ||\mathbf{q}_i[n]-\mathbf{g} ||_2, \forall i \in \{1,2\}, \forall n.$
We suppose that both \acp{uav} fly at   sufficiently high  altitudes to allow
obstacle-free communication with the \ac{gs} over a \ac{los} link. Thus, based on the free-space path loss model and \ac{fdma}, the
instantaneous throughput  from $U_i, \forall i\in \{1,2\},$ to the \ac{gs} is given by:
\begin{align}
 &R_{i,n}(\mathbf{q}_i,\mathbf{P}_{\mathrm{com},i})= B_{c,i} \; \log_2\left(1+\frac{P_{\mathrm{com},i}[n] \;\gamma}{d_{i,n}^2(\mathbf{q}_i)}\right), \forall n,
\end{align}
where  $B_{c,i}$ is the fixed communication bandwidth allocated for $U_i$ and $\gamma$ is the reference channel gain\footnote{The reference channel gain is the channel power gain at a reference distance of 1 m.} divided by the noise variance.

\section{Problem Formulation and Solution}
\subsection{Problem Formulation}
In this paper, we aim to maximize the  \ac{insar} coverage by jointly optimizing the \ac{uav} formation $\{\mathbf{q}_1,\mathbf{q}_2\}$ and the communication resources $\{\mathbf{P}_{\mathrm{com,1}},\mathbf{P}_{\mathrm{com,2}} \}$ while satisfying communication and interferometric quality-of-service constraints. To this end, the following optimization problem is formulated: 

\begin{alignat*}{2} 
&(\mathrm{P.1}):\max_{\mathbf{q}_1,\mathbf{q}_2, \mathbf{P}_{\mathrm{com,1}},\mathbf{P}_{\mathrm{com,2}}} \hspace{3mm}  C_N(\mathbf{q}_1,\mathbf{q}_2)   & \qquad&  \\
\text{s.t.} \hspace{3mm} &  \mathrm{C1: } \; z_{\mathrm{min}} \leq z_i \leq z_{\mathrm{ max}}, \forall i \in \{ 1,2\},               &      &  \\ & \mathrm{C2}: \;  x_1=x_t - z_1 \tan(\theta_d),             &      &  \\& \mathrm{C3}: \;  r_2(\mathbf{q}_2) \leq r_1(\mathbf{q}_1),             &      &  \\& \mathrm{C4}: \;    x_2 \leq x_t,             &      &  \\&  \mathrm{C5}:    b(\mathbf{q}_1,\mathbf{q}_2)  \geq b_{\mathrm{min}},  &      &     
 \\
  &  \mathrm{C6}:  \gamma_{\mathrm{SNR},n}(\mathbf{q}_1,\mathbf{q}_2)\geq \gamma_{\mathrm{SNR}}^{\mathrm{min}}, \forall n,        &      &     
 \\
 &  \mathrm{C7}: \gamma_{\mathrm{Rg}}(\mathbf{q}_2) \geq \gamma_{\mathrm{Rg}}^{\mathrm{min}},            &      &     
 \\
 &    \mathrm{C8}: \; h_{\mathrm{amb}}^{\mathrm{min}} \leq   h_{\mathrm{amb}}(\mathbf{q}_1,\mathbf{q}_2) \leq h_{\mathrm{amb}}^{\mathrm{max}},           &      & \\
& \mathrm{C9}: 0 \leq P_{\mathrm{com},i}[n]  \leq P_{\mathrm{com}}^{\mathrm{max}}, \forall \; i \in \{1,2\}, \forall n,    & &\\
& \mathrm{C10}: R_{i,n}(\mathbf{q}_i,\mathbf{P}_{\mathrm{com},i}) \geq R_{\mathrm{min},i}, \forall \; i \in \{1,2\}, \forall n,       & &  \\
& \mathrm{C11}:  \sum_{n=1}^{N} P_{\mathrm{com},i}[n] \delta_t \leq E_{\mathrm{com}} , \forall i \in \{1,2\}, \forall n.    & & 
\end{alignat*}
Note that the \ac{insar} coverage $C_N$ differs from the cooperative sensing coverage \cite{detection_probability}.
Constraint $\mathrm{ C1}$ define the maximum and minimum allowed flying altitude, denoted by $z_{\mathrm{max}}$ and $z_{\mathrm{min}}$,  respectively. Constraints $\mathrm{ C2}$ and $\mathrm{ C3}$ ensure maximum overlap between the beam footprint of the master drone and the area of interest. Constraint $\mathrm{ C4}$ is imposed because a side-looking \ac{sar} is assumed. Constraint $\mathrm{ C5}$ ensures safe operation, where $b_{\mathrm{min}}$ is the minimum separation distance of the two drones. Constraints $\mathrm{ C6}$ and $\mathrm{ C7}$ ensure minimum required coherence thresholds on the sensing $\mathrm{ SNR}$ and baseline decorrelation, denoted by $\gamma_{\mathrm{SNR}}^{\mathrm{min}}$ and $\gamma_{\mathrm{Rg}}^{\mathrm{min}}$,  respectively. Constraint $\mathrm{ C8}$ imposes minimum and maximum  \acp{hoa} denoted by $h_{\mathrm{amb}}^{\mathrm{min}}$ and $h_{\mathrm{amb}}^{\mathrm{max}}$, respectively, that satisfy prescribed \ac{dem} requirements \cite{coherence1}. Note that constraints $\mathrm{ C6}$, $\mathrm{ C7}$, and $\mathrm{ C8}$ are specific to \ac{insar} applications and have not been considered in existing \ac{uav}-based optimization frameworks. In fact, if the baseline of the \ac{uav} formation is too large, the signals are corrupted due to a high baseline decorrelation, while if the baseline is too small, the sensitivity to the ground topology is reduced \cite{tradeoff}. Constraint $\mathrm{ C9}$ ensures that the communication transmit power is non-negative and does not exceed the maximum allowed level denoted by $P_{\mathrm{com}}^{\mathrm{max}}$. Constraint $\mathrm{ C10}$ ensures that the achievable throughput of drone $U_i$ does not fall below the minimum required data rate $R_{\mathrm{min},i}$, which corresponds to an upper bound on the amount of  sensing data collected by $U_i$.  Constraint $\mathrm{ C11}$ limits the consumed communication energy to  $E_{\mathrm{ com}}$. Note that some constraints do not depend on time $n$ as some variables, such as $x_1$ and $z_2$, are optimized but are fixed across time due to the prescribed linear \ac{insar} trajectory imposed by the stripmap mode \ac{sar}  operation \cite{fixed_heading}. 
\subsection{Solution of the Optimization Problem}
Problem $\mathrm{(P.1)}$ is non-convex due to its objective function and constraints $\mathrm{C3}$ and $\mathrm{ C5-C8}$, which are non-convex and involve coupled optimization variables $\mathbf{q}_1$ and $\mathbf{q}_2$. In general, it is very challenging to find the globally optimal solution to problem $\mathrm{ (P.1)}$ and the available optimal algorithms suffer from high time complexity. To strike  a balance between performance and complexity,  we provide a low-complexity sub-optimal solution for the formulated problem based on \ac{ao}. To this end, problem $\mathrm{ (P.1)}$ is divided into two sub-problems, namely $\mathrm{(P.1.a)}$ and $\mathrm{(P.1.b)}$. 
\subsubsection{Slave \ac{uav} Optimization}
First, problem ($\mathrm{P.1}$) is solved for fixed  $\{ \mathbf{q}_1, \mathbf{P}_{\mathrm{com,1}}\}$. The resulting problem, denoted by sub-problem $\mathrm{(P.1.a)}$,
 is still non-convex and difficult to solve due to its objective function as well as non-convex constraints $\mathrm{ C5}$, $\mathrm{ C7}$, and $\mathrm{ C8}$. Yet, we provide a low-complexity sub-optimal solution based on \ac{sca}. In a first step, we replace the non-concave objective function with an equivalent concave function.\looseness=-1
  \begin{proposition}\label{prop:objective_function}
Non-concave objective function $C_N$ can be equivalently replaced by the following concave function: 
\ifonecolumn
\begin{align} \label{eq:objective_function_relaxation}
     \widetilde{C}(\mathbf{q}_1,\mathbf{q}_2)= \min(x_1 +\tan(\theta_{\mathrm{far}})z_1,x_2+ \tan(\theta_{\mathrm{far}})z_2)- \max(x_1 + \tan(\theta_{\mathrm{near}})z_1 ,x_2 + \tan(\theta_{\mathrm{near}}) z_2 ).
\end{align}
\else 

\begin{align} \label{eq:objective_function_relaxation}
       \widetilde{C}(\mathbf{q}_1,\mathbf{q}_2)&= \min(x_1 +\tan(\theta_{\mathrm{far}})z_1,x_2+ \tan(\theta_{\mathrm{far}})z_2)- \notag \\ & \max(x_1 + \tan(\theta_{\mathrm{near}})z_1 ,x_2 + \tan(\theta_{\mathrm{near}}) z_2 ),
\end{align}
\fi 
\end{proposition}	
\ifarxiv

\begin{proof}
The objective function of problem $\mathrm{(P.1)}$, denoted by $C_N(\mathbf{q}_1,\mathbf{q}_2)$, can be expressed as  $\delta_t \sum\limits_{n=1}^{N}v_y[n]\max(0,\widetilde{C}(\mathbf{q}_1,\mathbf{q}_2))$. Therefore, for all $ \mathbf{q}_1$ and $ \mathbf{q}_2$, $ C_N(\mathbf{q}_1,\mathbf{q}_2)\geq  \delta_t \min\limits_n(v_y[n])\widetilde{C}(\mathbf{q}_1,\mathbf{q}_2)$ holds, meaning that maximizing $\widetilde{C}$ results in maximizing $C_N$. Next, we show that the optimal values for $\widetilde{C}$ and $C_N$ are related by the proportionality constant $\delta_t \sum\limits_{n=1}^{N}v_y[n]$. Let the optimal bistatic formation for problem $\mathrm{(P.1)}$ be denoted by   $\{\mathbf{q}^*_1,\mathbf{q}^*_2\}$. Here, there are two possible outcomes; (i) if                 $C_N(\mathbf{q}^*_1,\mathbf{q}^*_2)= 0$, i.e., there is no feasible   solution  with non-negative \ac{insar} coverage, then optimizing $\widetilde{C}$ results in reducing the distance along the $x$-axis between the ground  footprints of the master and slave \acp{uav}, which is not relevant to problem $\mathrm{(P.1)}$ as there is no overlap that can be used for interferometry, (ii) if $C_N(\mathbf{q}^*_1,\mathbf{q}^*_2)> 0$, then it is easy to note that  $C_N(\mathbf{q}^*_1,\mathbf{q}^*_2)=\delta_t \sum\limits_{n=1}^{N}v_y[n]\widetilde{C}(\mathbf{q}^*_1,\mathbf{q}^*_2)$. The outcome of (i) and (ii) proves that objective functions $C_N$ and  $\widetilde{C}$ are equivalent.
\end{proof}
\fi
Based on \textbf{Proposition} \ref{prop:objective_function}, we equivalently maximize objective function $ \widetilde{C}$  instead of $C_N$. Next, we approximate non-convex constraint $\mathrm{C5}$ with a convex constraint. 
\begin{lem} \label{lem:C5}
A first-order convex approximation for the concave term $-b^2(\mathbf{q}_1,\mathbf{q}_2)$ around an arbitrary point $\mathbf{a} \in \mathbb{R}^2$ and for fixed $\mathbf{q}_1$ is obtained based on surrogate functions as follows: 
\begin{align}
  g(\mathbf{q}_2)= b^2(\mathbf{q}_1,\mathbf{q}_2)-  2 (\mathbf{a} -\mathbf{q}_1)^T(2\mathbf{q}_2 -\mathbf{a}-\mathbf{q}_1).\label{eq:g-f}
\end{align}
\end{lem}
\ifarxiv
	\begin{proof} 
		In this proof, we check the first-order conditions for the proposed surrogate function denoted by $g$, which are majorization and smoothness\cite{surrogate}. First, let $f(\mathbf{q}_2)=-b^2(\mathbf{q}_1,\mathbf{q}_2)$. The gradient of the difference $g-f$ \ac{wrt} $\mathbf{q}_2$ is given by: 
		\begin{align}
			\nabla_{\mathbf{q}_2 } (g-f)(\mathbf{q}_2)= 4 (\mathbf{q}_2 -\mathbf{q}_1 )-4(\mathbf{a} -\mathbf{q}_1).\label{eq:gradient_g-f}
		\end{align}
		Based on (\ref{eq:g-f}) and (\ref{eq:gradient_g-f}),  we can easily show that the difference $g-f$ is non-negative with $(g-f)(\mathbf{a})=0$, which proves majorization. Second, the difference $g-f$ is differentiable, its gradient is continuous, and $  \nabla_{\mathbf{q}_2 } (g-f)(\mathbf{a})=0$, which proves smoothness and concludes the proof.
\end{proof}
\fi
Based on \textbf{Lemma} \ref{lem:C5}, in the $j^{\mathrm{th}}$ iteration of the \ac{sca} algorithm, constraint $\mathrm{ C5}$ can be approximated around point $\mathbf{q}^{(j)}_2 \in \mathbb{R}^2$ by the following convex constraint:
\begin{align}
{\mathrm{\widetilde{C5}}:    b^2(\mathbf{q}_1,\mathbf{q}_2)  -  2 (\mathbf{q}^{(j)}_2 -\mathbf{q}_1)^T(2\mathbf{q}_2-\mathbf{q}^{(j)}_2-\mathbf{q}_1)\leq -b_{\mathrm{min}}^2.}\label{eq:C5}
\end{align}

Next, we use first-order Taylor expansion around point $\mathbf{q}_2^{(j)}\in \mathbb{R}^2$ to provide a convex approximation for constraint $\mathrm{ C7}$:
\begin{align}
    \widetilde{\mathrm{ C7}}:& (x_2-x_t)-A \sin(\theta_1)r_2(\mathbf{q}_2^{(j)})- \notag \\ &A \sin(\theta_1)\nabla_{\mathbf{q}_2}r_2(\mathbf{q}_2^{(j)})^T(\mathbf{q}_2-\mathbf{q}_2^{(j)})\leq 0,
\end{align}
where $A=\frac{-\gamma_{\mathrm{ Rg}}^{\mathrm{min}}B_{ p}- 2+B_{p} }{\gamma_{\mathrm{ Rg}}^{\mathrm{min}}B_{ p}- 2-B_{ p}}\geq 0$.

\begin{proposition} \label{prop:C8}
Based on  $\mathrm{C2}$ and $\mathrm{C3}$,  the perpendicular baseline is independent of $\mathbf{q}_1$, and can be rewritten as: 
\begin{gather}
     b_{\bot}(\mathbf{q}_2)=\frac{1}{\sqrt{\tan(\theta_1)^2+1}}\Big|(x_t-x_2)-\tan(\theta_1)z_2\Big|.   \label{ap:perpendicular_baseline}
\end{gather}
\end{proposition}
\ifarxiv 
\begin{proof}
		Please refer to Appendix. A.
\end{proof}
\fi 
Based on Taylor expansion and \textbf{Proposition} \ref{prop:C8}, non-convex constraint $\mathrm{ C8}$ can be approximated around point $\mathbf{q}_2^{(j)}\in \mathbb{R}^2$ by the following convex constraints: 
\begin{gather}
    \widetilde{\mathrm{C8a: }} ((x_t-x_2)-\tan(\theta_1)z_2)^2\leq \frac{a}{{\left(h_{\mathrm{amb}}^{\mathrm{min}}\right)}^2}, \\
    \mathrm{\widetilde{C8b}: } J(\mathbf{q}_2^{(j)})+ \nabla_{\mathbf{q}_2}J(\mathbf{q}_2^{(j)})^T (\mathbf{q}_2 -\mathbf{q}_2^{(j)} )\leq \frac{-a}{{ \left(h_{\mathrm{amb}}^{\mathrm{max}}\right)}^2},
\end{gather}
where $a=(\tan^2(\theta_1)+1)\lambda c \tan(\theta_1) z_1$ and $J(\mathbf{q}_2)=-((x_t-x_2)-\tan(\theta_1)z_2)^2$ .
To summarize, sub-problem $\mathrm{ (P.1.a)}$ is approximated by the following convex optimization problem: 
\begin{alignat*}{2} 
&(\widetilde{\mathrm{P.1.a}}):\max_{\mathbf{q}_2,\mathbf{P}_{\mathrm{com,2}}} \hspace{3mm}  \widetilde{C}(\mathbf{q}_1,\mathbf{q}_2)   & \qquad&  \\
&\text{s.t.} \hspace{3mm} \mathrm{C1 - C4, \widetilde{C5},C6, \widetilde{C7},\widetilde{ C8a}, \widetilde{C8b}, C9-C11.}  &   & 
\end{alignat*}

The proposed procedure to solve sub-problem $\mathrm{(P.1.a)}$ is  summarized in \textbf{Algorithm} \ref{euclid}, where the convex approximation  $\mathrm{ (\widetilde{P.1.a})}$ is solved  using the Python convex optimization library CVXPY \cite{cvxpy}.  \textbf{Algorithm} \ref{euclid} converges to a local optimum  of sub-problem $\mathrm{(P.1.a)}$ in polynomial time complexity \cite{sca_complexity}. 
\begin{algorithm}
  \caption{Successive Convex Approximation for $\mathrm{ (P.1.a)}$ }\label{euclid}
  \begin{algorithmic}[1] 
        \label{algorithm1} \State For fixed $\{\mathbf{q}_1,\mathbf{P}_{\mathrm{ com,1}}\}$, set initial point $\{\mathbf{q}_2^{(1)},\mathbf{P}_{\mathrm{ com,2}}^{(1)}\}$, iteration index $j=1$, and error tolerance $0< \epsilon \ll 1$.
      \State \textbf{repeat}
             \State $  $Determine coverage $ \widetilde{C}(\mathbf{q}_1,\mathbf{q}_2), \hspace{1mm} \mathbf{q}_2, $ and $ \mathbf{P}_{\mathrm{ com,2}} $ by solving $\mathrm{ \widetilde{(P.1.a)}}$ around point $\{\mathbf{q}_2^{(j)},{\mathbf{P}^{(j)}_{\mathrm{ com,2}}}\}$.           \State$ $Set $ j=j+1,\hspace{1mm} \mathbf{q}_2^{(j)}= \mathbf{q}_2,\hspace{1mm} {\mathbf{P}^{(j)}_{\mathrm{ com,2}}}=\mathbf{P}_{\mathrm{ com,2}}.$ 
            \State \textbf{until} $\big |\frac{\widetilde{C}(\mathbf{q}_1,\mathbf{q}_2^{(j)})-\widetilde{C}(\mathbf{q}_1,\mathbf{q}_2^{(j-1)})}{\widetilde{C}(\mathbf{q}_1,\mathbf{q}_2^{(j)})}\big|\leq \epsilon$
       \State \textbf{return} solution
  \end{algorithmic}
\end{algorithm}
\subsubsection{Master \ac{uav} Optimization}
Next, problem ($\mathrm{P.1}$) is solved for fixed  $\{ \mathbf{q}_2, \mathbf{P}_{\mathrm{com,2}}\}$. The resulting problem, denoted by sub-problem $\mathrm{(P.1.b)}$,
 is still non-convex due to the objective function and constraints $\mathrm{ C5}$ and $\mathrm{ C8}$. Problem $(\mathrm{P.1.b})$  is solved based on  \ac{sca}, similar to $(\mathrm{P.1.a})$. First, we replace the objective function $C_N$ with $\widetilde{C}$ as in (\ref{eq:objective_function_relaxation}). Similar to (\ref{eq:C5}), and based on \textbf{Lemma} \ref{lem:C5}, in the $j^{\mathrm{th}}$ iteration of the \ac{sca} algorithm, constraint $\rm C5$ is approximated around point $ \mathbf{q}^{(j)}_1 \in \mathbb{R}^2 $: 
\begin{align}
 {\rm \widetilde{\widetilde{C5}}:  b^2(\mathbf{q}_1,\mathbf{q}_2)  -  2 (\mathbf{q}^{(j)}_1 -\mathbf{q}_2)^T(2\mathbf{q}_1 -\mathbf{q}^{(j)}_1-\mathbf{q}_2)\leq -b_{\mathrm{min}}^2.}
\end{align}
Based on \textbf{Proposition} \ref{prop:C8}
 and first-order Taylor expansion, constraint $\rm C8$ is approximated around point $ \mathbf{q}^{(j)}_1 \in \mathbb{R}^2 $ by the following convex constraints: 
\begin{align}
\widetilde{\widetilde{\mathrm{C8a}}}: r_1 (\mathbf{q}_1) \leq \frac{h^{\mathrm{max}}_{\mathrm{ amb}}  b_{\bot}(\mathbf{q}_2)}{\lambda \sin(\theta_1)}, 
\end{align}
\begin{align}
\mathrm{\widetilde{\widetilde{C8b}}}: r_1^2(\mathbf{q}_1^{(j)}) +\nabla_{\mathbf{q}_1} r_1(\mathbf{q}_1^{(j)})^T (\mathbf{q}_1-\mathbf{q}_1^{(j)}) \geq  \Big(\frac{h^{\mathrm{min}}_{\mathrm{ amb}}  b_{\bot}(\mathbf{q}_2)}{\lambda \sin(\theta_1)}\Big)^2.
\end{align}
To summarize, sub-problem $\rm (P.1.b)$ is approximated by the following convex optimization problem: 
\begin{alignat*}{2} 
&(\widetilde{\mathrm{P.1.b}}):\max_{\mathbf{q}_1,\mathbf{P}_{\mathrm{com,1}}} \hspace{3mm}  \widetilde{C}(\mathbf{q}_1,\mathbf{q}_2)   & \qquad&  \\
&\text{s.t.} \hspace{3mm} \mathrm{C1-C3, \widetilde{\widetilde{C5}}, C6,\widetilde{\widetilde{C8a}},\widetilde{\widetilde{C8b}},C9-C11.}  &   & 
\end{alignat*}

Similar to $\mathrm{ (P.1.a)}$, problem  $\mathrm{ (P.1.b)}$ is solved based on \ac{sca}, where the convex approximation  $\mathrm{ \widetilde{(P.1.b)}}$ is solved using CVXPY \cite{cvxpy}.  The algorithm converges to a local optimum in polynomial time complexity \cite{sca_complexity}. As the proposed algorithm is similar to \textbf{Algorithm} \ref{euclid}, its detailed steps are omitted.

\subsection{Solution to Problem $\rm (P.1)$}
To summarize, to solve problem $\rm (P.1)$, we use \ac{ao} by solving sub-problems $\rm (P.1.a)$ and  $\rm (P.1.b)$ iteratively.  In \textbf{Algorithm} \ref{alg:ao}, we summarize all the steps of the solution to problem $\mathrm{(P.1)}$. Based on \cite{notes_ao}, \textbf{Algorithm} \ref{alg:ao} converges to a local optimum of problem  ($\mathrm{P.1}$) in polynomial time complexity. In practice, this outcome is achieved in just a few iterations. 

\begin{algorithm}[]
	\caption{Alternating Optimization Algorithm}\label{alg:ao}
	\begin{algorithmic}[1] 
		\State Set initial formation $(\mathbf{q}_1^{(1)},\mathbf{q}_2^{(1)})$, initial communication resources $(\mathbf{P}^{(1)}_{\mathrm{com,1}},\mathbf{P}^{(1)}_{\mathrm{com,2}} )$, iteration index $k=1$, and error tolerance $0< \epsilon \ll 1$.
		\State \textbf{repeat}
		\State $  $ Determine coverage $ \widetilde{C}(\mathbf{q}_1^{(k)},\mathbf{q}_2), \hspace{1mm} \mathbf{q}_2,$ and $ \mathbf{P}_{\mathrm{com,2}}$ by solving $\rm (P.1.a)$ for fixed $(\mathbf{q}_1^{(k)},{\mathbf{P}^{(k)}_{\mathrm{com,1}}})$ with initial point $(\mathbf{q}_2^{(k)},\mathbf{P}^{(k)}_{\mathrm{com,2}} )$ using \textbf{Algorithm} \ref{euclid}.    
		\State $ $ Set $k=k+1$,  $\mathbf{q}_2^{(k)}=\mathbf{q}_2 $ and $\mathbf{P}^{(k)}_{\mathrm{com,2}} =\mathbf{P}_{\mathrm{com,2}}. $
		\State $  $ Determine coverage $ \widetilde{C}(\mathbf{q}_1,\mathbf{q}_2^{(k)}), \hspace{1mm} \mathbf{q}_1,$ and $ \mathbf{P}_{\mathrm{com,1}}$ by solving $\rm (P.1.b)$ for fixed $(\mathbf{q}_2^{(k)},{\mathbf{P}^{(k)}_{\mathrm{com,2}}})$ with initial point $(\mathbf{q}_1^{(k-1)},\mathbf{P}^{(k-1)}_{\mathrm{com,1}} )$.    
		\State $ $ Set $\mathbf{q}_1^{(k)}=\mathbf{q}_1, $ and ${\mathbf{P}^{(k)}_{\mathrm{com,1}}}=\mathbf{P}_{\mathrm{com,1}}. $
		\State \textbf{until} $\big |\frac{ \widetilde{C}(\mathbf{q}_1^{(k)},\mathbf{q}_2^{(k)})-\widetilde{C}(\mathbf{q}_1^{(k-1)},\mathbf{q}_2^{(k-1)})}{\widetilde{C}(\mathbf{q}_1^{(k)},\mathbf{q}_2^{(k)})}\big|\leq \epsilon$
		\State \textbf{return} solution $\{\mathbf{q}_1^{(k)}, \mathbf{q}_2^{(k)},\mathbf{P}^{(k)}_{\mathrm{com,1}},\mathbf{P}^{(k)}_{\mathrm{com},2} \}$
	\end{algorithmic}
\end{algorithm}
\begin{table}[]
\centering
\caption{System parameters \cite{victor,14,snr_equation,coherence1}.}
\label{tab:my-table}
\begin{tabular}{|c|c?c|c?c|c|}
\hline
Parameter           & Value & Parameter & Value & Parameter &value \\ \hline
$z_{\mathrm{min}}$&  1 m     &$x_g=y_g $        &-93 m            &$f_0$    & 2.5 GHz \\ \hline
$z_{\mathrm{max}}$&  100 m   &$z_g$             &2 m           &$B_{\mathrm{Rg}}$    &3 GHz  \\ \hline
$h_{\mathrm{amb}}^{\mathrm{min}}$&0.6 m     &$P_{\mathrm{com}}^{\mathrm{max}}$&10 dB &$T_{\mathrm{sys}}$   & 400 K \\ \hline
$h_{\mathrm{amb}}^{\mathrm{max}}$&2 m      &$R_{\mathrm{ min},i}$   &1 Mbits&$L_{\mathrm{sys}}$    & 2 dB \\ \hline
$x_t$             &20 m      &$E_{\mathrm{com}}$&700 J       &$L_{\mathrm{azm}}$    & 2 dB \\ \hline
$b_{\mathrm{min}}$&2 m    	 &$B_{c,i}$             & 1 GHz             &$L_{\mathrm{atm}}$    & 0 dB \\ \hline
$\delta_t$ 		  &0.5 s       &$\gamma$          &20 dB          &$F$    & 5 dB\\ \hline
$N$               &$10^2$   &$\theta_{3\mathrm{dB}}$&30°  &$\sigma_0$    &-5 dBm$^2$ \\ \hline
$\gamma_{\mathrm{Rg}}^{\mathrm{min}}$        &0.8     &$\theta_d$        & 45°         &$\tau_p\times\mathrm{PRF}$    & 0.8 \\ \hline
$\gamma_{\mathrm{SNR}}^{\mathrm{min}}$&$0.8$ &$v_y$&  2 m/s       & $G_t$   & 6 dBi  \\ \hline
$P_t$		&$15$ dBm	&$\lambda$&  0.12 m       & $G_r$   & 6 dBi  \\ \hline
\end{tabular}
\end{table}

\section{Simulation Results and Discussion }
In this section, we present simulation results for the proposed \ac{uav} formation and resource allocation algorithm. The system parameters are specified in Table \ref{tab:my-table}. The proposed solution is compared with the two following benchmark schemes: \\
\textbf{Benchmark scheme 1:} Here, we employ a vertical bistatic formation, i.e., we optimize the $x$-position of one of the drones and impose $x_1 = x_2$ \cite{dinsar}. The remaining optimization variables are determined based on \textbf{Algorithm} \ref{alg:ao}.\\
\textbf{Benchmark scheme 2:} Here, we use equal communication powers, i.e., we optimize $\mathbf{P}_{\mathrm{com},i}[1]$ and enforce  $\mathbf{P}_{\mathrm{com},i}[n] =\mathbf{P}_{\mathrm{com},i}[n-1]$, $\forall i \in \{1,2\}$, $\forall n\geq2$. The remaining optimization variables are determined by \textbf{Algorithm} \ref{alg:ao}. \par

\ifonecolumn
\begin{figure}
\centering
    \includegraphics[width=4 in]{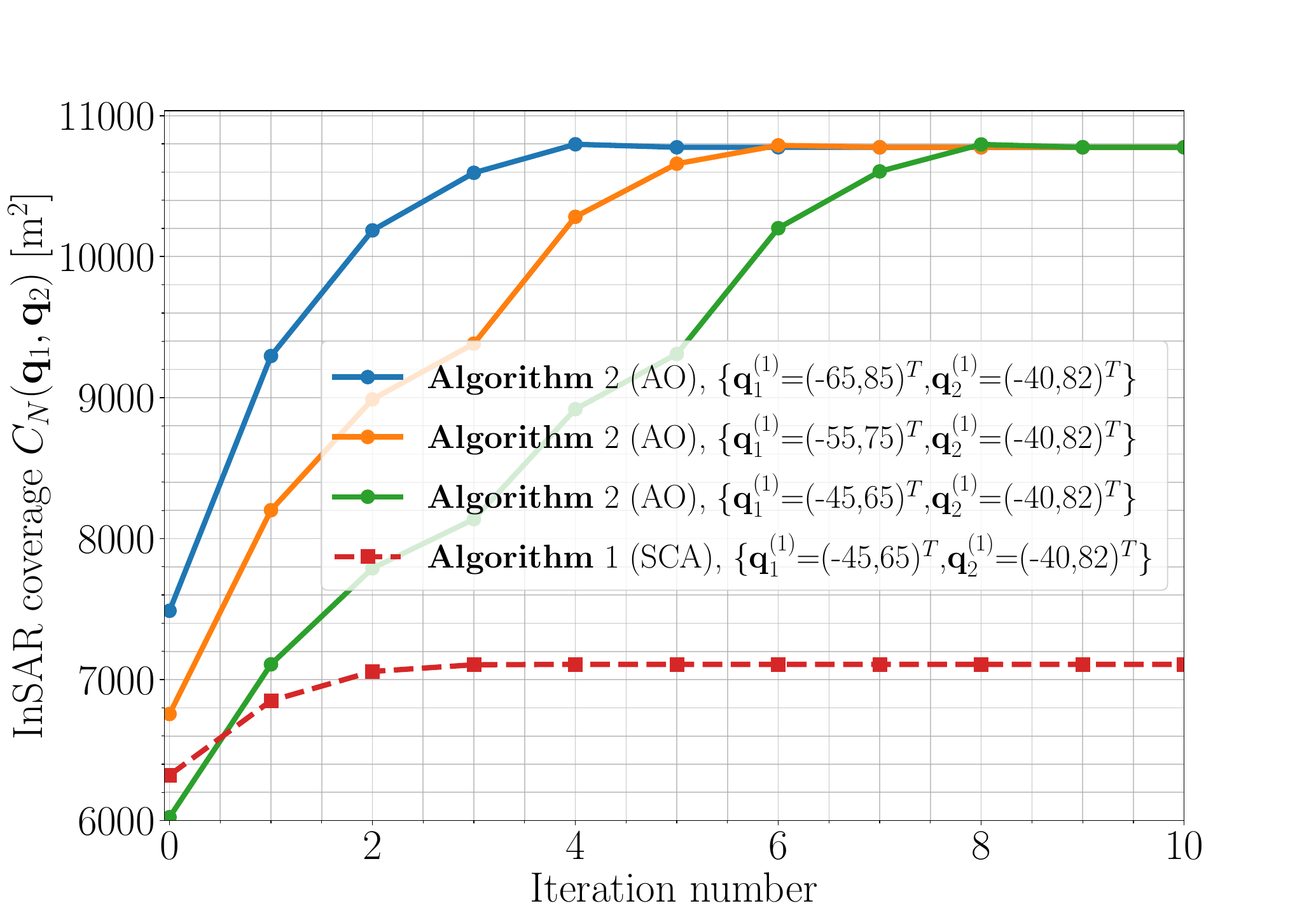}
    \caption{Convergence of  the proposed solution for different initial bistatic UAV formations. The error tolerance is $\epsilon=10^{-4}$.}
    \label{fig:convergence}
\end{figure}
\else
\begin{figure}
\centering
    \includegraphics[width=1\columnwidth]{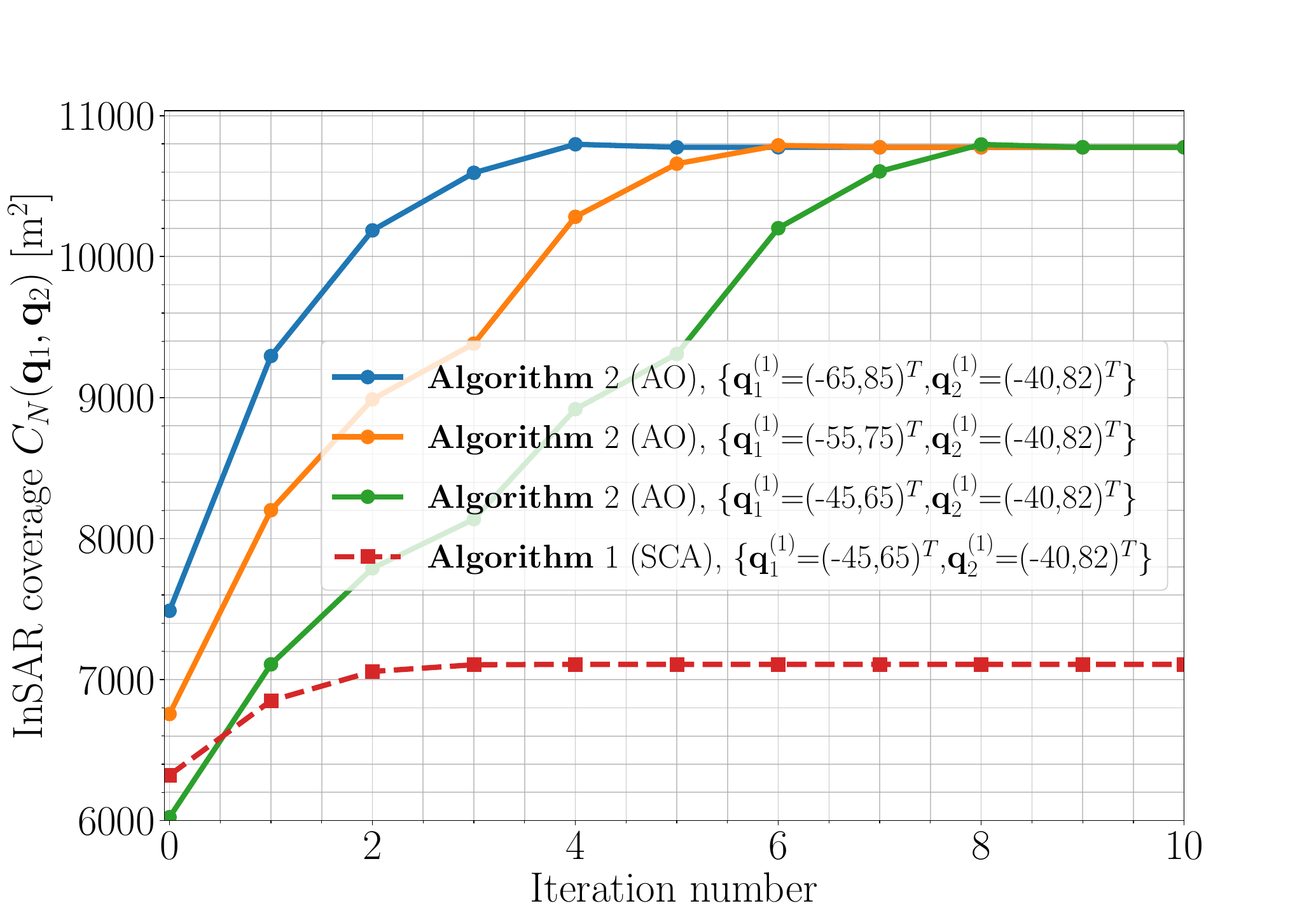}
    \caption{Convergence of  the proposed solution for different initial bistatic UAV formations. The error tolerance is $\epsilon=10^{-4}$.}
    \label{fig:convergence}
\end{figure}
\fi

Figure \ref{fig:convergence} illustrates the convergence of the proposed \textbf{Algorithm} \ref{alg:ao}, which solves problem $\mathrm{(P.1)}$, as well as  that of \textbf{Algorithm}  \ref{euclid}, which solves sub-problem $\mathrm{(P.1.a)}$.   For different initial bistatic formations, \textbf{Algorithm} \ref{alg:ao} converges consistently to the same objective function value. Though the convergence rate depend on the initial \ac{uav} formation, the optimal value is found in a few iterations for both algorithms. \par
\ifonecolumn
\begin{figure}[h]
	\centering
	\includegraphics[width=4 in]{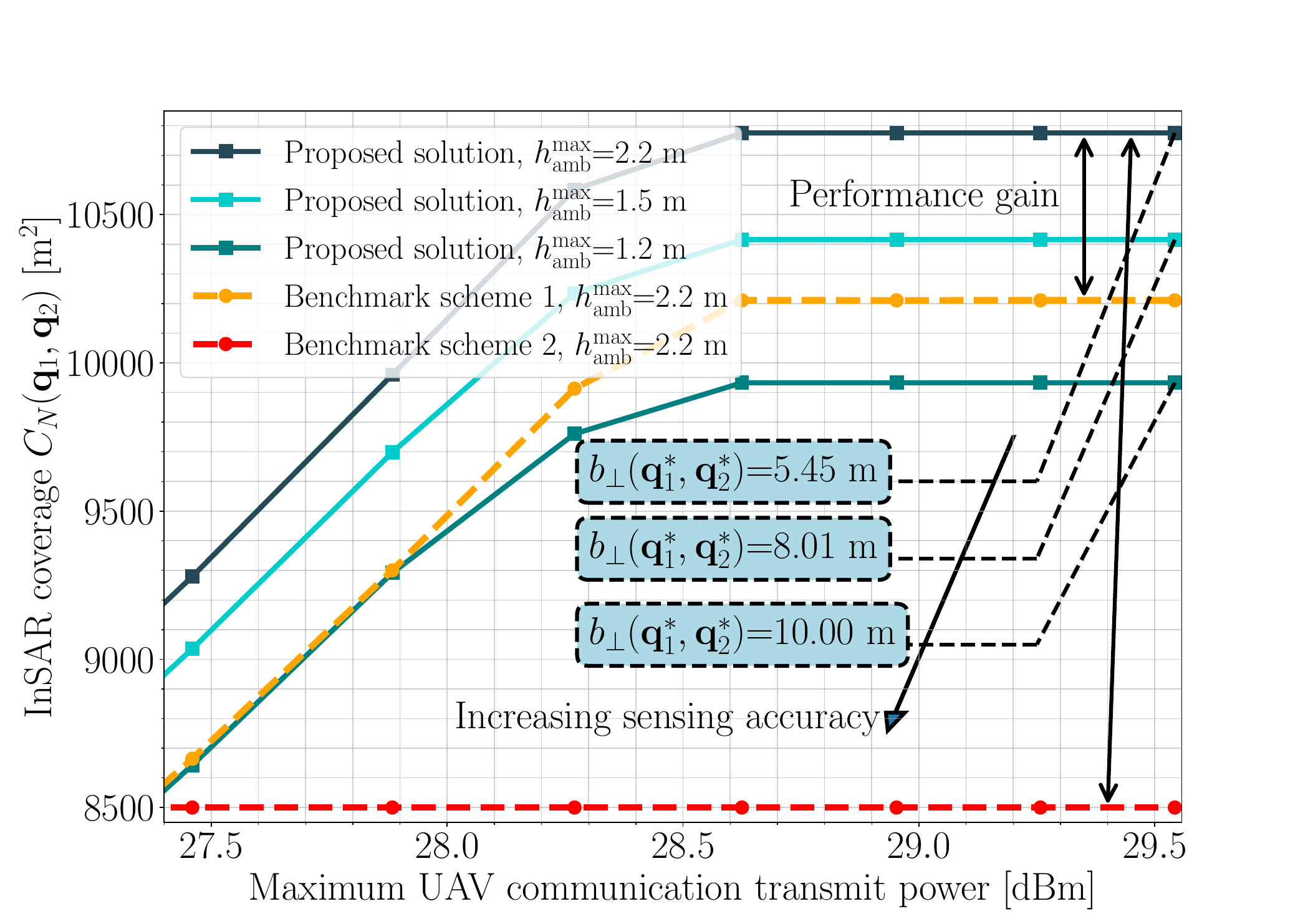}
	\caption{InSAR total coverage versus maximum UAV communication transmit power, $P_{\mathrm{com}}^{\mathrm{max}}$.}
	\label{fig:Maximum_Power}
	
\end{figure}
\else
\begin{figure}
	\centering
	\includegraphics[width=1\columnwidth]{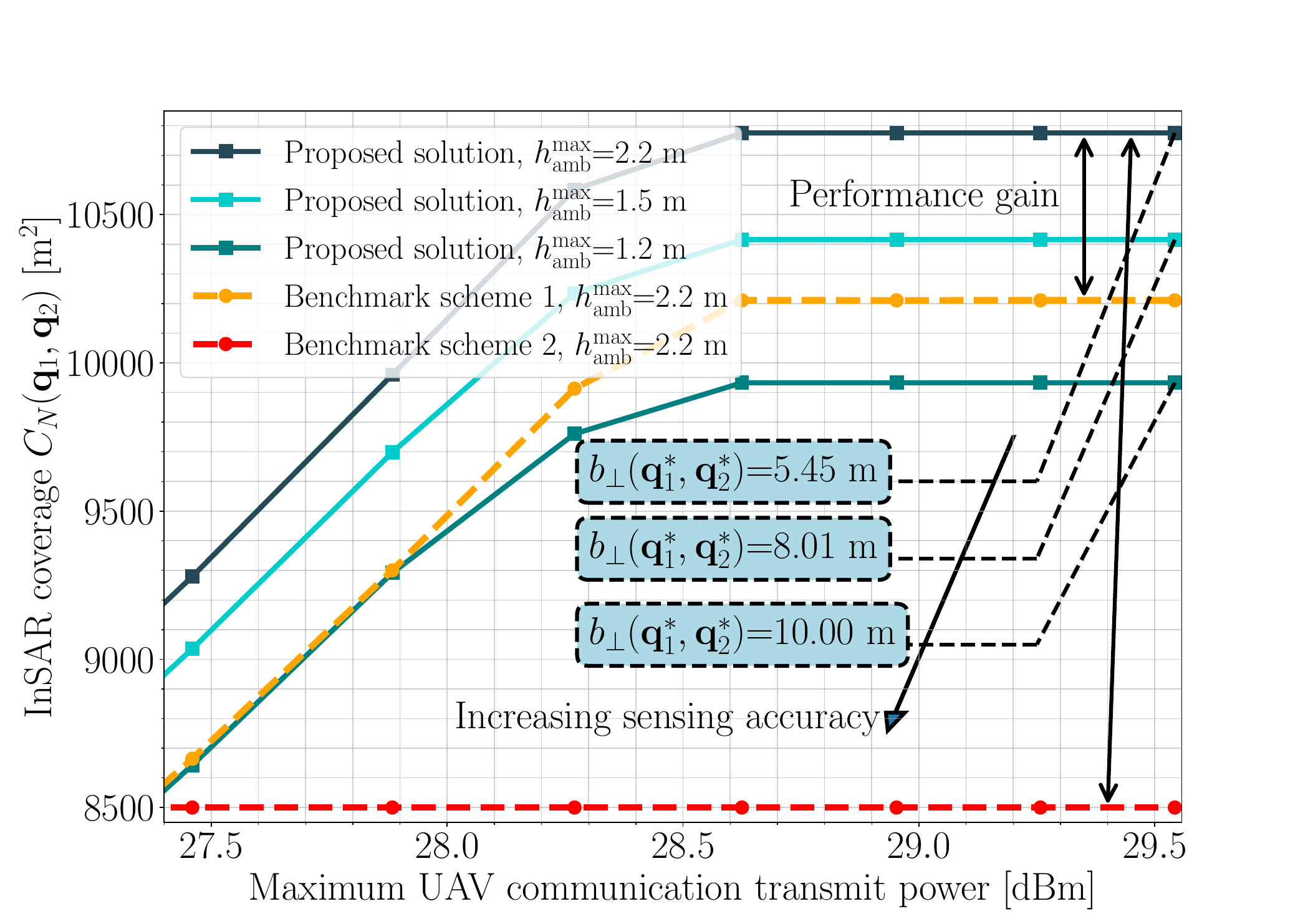}
	\caption{InSAR total coverage versus maximum UAV communication transmit power, $P_{\mathrm{com}}^{\mathrm{max}}$.}
	\label{fig:Maximum_Power}
\end{figure}
\fi

Figure \ref{fig:Maximum_Power} shows the achieved \ac{insar} coverage versus the maximum \ac{uav} communication power, $P_{\mathrm{com}}^{\mathrm{max}}$. Benchmark scheme 2, which employs a static communication power allocation, exhibits the lowest performance. Benchmark scheme 1 achieves a higher performance than benchmark scheme 2, due to the optimization of the communication power allocation, which leads to an enhanced range of the drones. For a maximum \ac{hoa} of $h^{\mathrm{max}}_{\mathrm{amb}}=2.2 $ m, the proposed solution substantially outperforms  benchmark schemes 1 and 2 as it jointly optimizes the \acp{uav}  formation and the communication resource allocation, with an additionally covered area of more than $575$ m$^2$ and $2300$ m$^2$, respectively. Figure \ref{fig:Maximum_Power} also reveals an interesting \ac{insar} performance trade-off; while a smaller \ac{hoa} improves the accuracy of the \ac{insar} sensing \cite{coherence1,insar_introduction,tradeoff}, it results in a reduced \ac{insar} coverage. This is because a lower \ac{hoa}  necessitates  a \ac{uav} formation with a larger perpendicular baseline, which in turn leads to an increased offset between the ground footprints of the master and slave radar systems. For instance, a maximum \ac{hoa} of 2.2 m requires a perpendicular baseline of only  5.54 m, whereas a maximum \ac{hoa} of 1.2 m  requires a perpendicular baseline of 10 m leading to a loss of $7.48\%$ in  total coverage.\looseness=-1  \par

\section{Conclusion}
In this paper, we investigated the \ac{uav} joint formation and communication resource allocation optimization for bistatic \ac{uav}-based \ac{insar} sensing, where \ac{insar} coverage, interferometric coherence, and \ac{hoa} were introduced as relevant \ac{insar} sensing performance metrics. A non-convex optimization problem was  formulated and solved for the maximization of the bistatic \ac{insar} ground coverage while enforcing the pertinent communication and sensing performance limits. Simulation results confirmed the  superior performance of the proposed algorithm compared to two  benchmark schemes and emphasized the important role of efficient communication resource allocation for maximum \ac{insar} coverage. We showed that \ac{uav} formations with a long perpendicular interferometric baseline, which achieve higher sensing accuracy, result in reduced \ac{insar} coverage. This creates an interesting sensing performance trade-off specific to \ac{uav}-based \ac{insar} systems.

\bibliographystyle{IEEEtran}
\bibliography{biblio}
\ifarxiv
	\section*{Appendix A: Proof of Proposition. \ref{prop:C8} }\label{appendix}
	
	In this appendix, we derive a simplified expression for the perpendicular baseline. It can be shown that the perpendicular baseline is fixed across the along-track direction, i.e., the $y$-direction. Henceforth, we only focus on the across-track plane where the coordinates of an arbitrary point are denoted by $(x,z)$. Based on constraints $\mathrm{C2}$ and $\mathrm{C3}$, the perpendicular baseline is the projection of the slave drone $U_2$ on $U_1$'s  \ac{los}. $U_1$'s  \ac{los}, i.e., the line passing through points $(x_t,0)$ and $(x_1,z_1)$, is characterized  by: 
	\begin{align}
		z=\tan(\theta_1)(x_t-x). \label{ap:line1}
	\end{align}
	The equation of the line perpendicular to (\ref{ap:line1}) passing through point $(x_2,z_2)$ is given by: 
	\begin{align}
		z-z_2=\tan(\theta_1)(x-x_2). \label{ap:line2}
	\end{align}
	Let point $p$ with coordinates $(x_p,z_p)$ be the intersection point between lines (\ref{ap:line1}) and (\ref{ap:line2}). The coordinates of $p$  are derived by solving the system of equations \{(\ref{ap:line1}),(\ref{ap:line2})\}, which leads to: 
	\begin{align}
	x_p&= \frac{\tan(\theta_1)}{\tan^2(\theta_1)+1}\left(\tan(\theta_1)x_2+\frac{1}{\tan(\theta_1)}x_t-z_2\right),\label{eq:projection1} \\
			z_p&=\frac{1}{\tan^2(\theta_1)+1}\left(\tan(\theta_1)(x_t-x_2)+z_2\right). \label{eq:projection2} 
	\end{align}
	Using the results in (\ref{eq:projection1}) and (\ref{eq:projection2}), and after some algebraic manipulations, we can show that  the perpendicular baseline, i.e., the distance between points  $(x_p,z_p)$ and $(x_2,z_2)$, is given by:
	\begin{gather}
		b_{\bot}(\mathbf{q}_2)=\frac{1}{\sqrt{\tan(\theta_1)^2+1}}\Big|(x_t-x_2)-\tan(\theta_1)z_2\Big|. \label{ap:perpendicular_baseline}
	\end{gather}
\fi
\end{document}